\documentclass[english,twocolumn,prl,showpacs]{revtex4}
\usepackage[latin1]{inputenc}
\usepackage{amsmath}
\usepackage{graphicx}
\usepackage[english]{babel}

\batchmode

\newcommand{\beq}{\begin{equation}}
\newcommand{\eeq}{\end{equation}}
\newcommand{\ket}[1]{| #1 \rangle}
\newcommand{\bra}[1]{\langle #1 |}

\begin{document}

\title{Monitoring stimulated emission at the single photon level in one-dimensional atoms}

\author{D. Valente$^{1}$}\email{daniel.valente@grenoble.cnrs.fr}
\author{S. Portolan$^{1}$}
\author{G. Nogues$^{1}$}
\author{J.P. Poizat$^{1}$}
\author{M. Richard$^{1}$}
\author{J.M. G\' erard$^{2}$}
\author{M. F. Santos $^{3}$}
\author{A. Auff\`eves$^{1}$}


\affiliation{$^{1}$ Institut N\'eel-CNRS, Grenoble, France}

\affiliation{$^{2}$ CEA/INAC/SP2M, Grenoble, France}

\affiliation{$^{3}$ Universidade Federal de Minas Gerais, Belo Horizonte, Brazil}

\begin{abstract}
We theoretically investigate signatures of stimulated emission at the single photon
level for a two-level atom interacting with a one-dimensional light field. We consider the transient
regime where the atom is initially excited, and the steady state regime where the atom is continuously
driven with an external pump. The influence of pure dephasing is studied, clearly showing that these effects
can be evidenced with state of the art solid state devices. We finally
propose a scheme to demonstrate the stimulation of one optical transition by monitoring another one,
in three-level one-dimensional atoms.
\end{abstract}
\pacs{42.50.Ct, 42.50.Gy}

\maketitle

Exploring light-matter interaction at the single photon level is a quest of quantum optics, that has been successfully achieved
so far with emitters in high quality factor microwave \cite{cqed} or optical cavities \cite{kimble}. High atom-field couplings are obtained at the price
of keeping the photons trapped in the mode, which may limit their exploitation for all practical purposes.  Alternative strategies have thus emerged, based on
the coupling of the emitter to a one-dimensional (1D) electromagnetic environment. A pioneering realization of such a ``1D atom'' consisted in an atom coupled
to a leaky directional cavity \cite{1D}. Nowadays, 1D atoms can be implemented in a wide range of physical systems, from
quantum dots embedded in photonic wires \cite{jmg}, in photonic crystal \cite{englund} or in plasmonic waveguides \cite{Lukin}, superconducting qubits in circuit QED \cite{circuitqed, astafiev},
to atoms \cite{atoms} and molecules in tightly focused beams \cite{sandoghdar}.
When probed with a resonant field, the natural directionality of 1D atoms allows to reach a high mode matching between the incoming and the scattered
light, manifested by the destructive interference of the two fields \cite{GNL,astafiev,englund,  sandoghdar}. Equivalently, perfect mode matching allows to saturate the emitter with a single photon \cite{GNL},
so that 1D atoms have been identified as promising single photon transistors \cite{Lukin} and two-photon gates \cite{kojima}.

This giant non-linear behavior strongly motivates to study the properties of the system when the atomic population is inverted,
and to revisit in the one-dimensional geometry the concept of stimulated emission introduced by Einstein \cite{einstein}.
Searching for signatures of stimulation at the single photon level not only provides new insights into a fundamental concept
of quantum optics, but also allows to envision appealing applications in quantum information processing. Optimal quantum cloning machines and
single photon adders could be implemented in these systems, which offer promising alternatives to cavity quantum electrodynamics
based devices, where these functionalities have been probed so far \cite{simon,maser}.
In this paper we theoretically characterize stimulation by single photons in two different regimes, namely the transient regime where the atom is initially
excited, and the steady-state one where the emitter is continuously excited by an incoherent light source. Signatures of stimulated emission are searched in the atomic population, and in the light field radiated by the atom.
To study the potential of solid state systems to demonstrate such effects, pure dephasing is taken into account. Finally,
the possibility of exploiting an ancillary atomic transition to monitor the stimulation is explored.

\begin{figure}[h!]
\begin{center}
\includegraphics[width=0.6\linewidth]{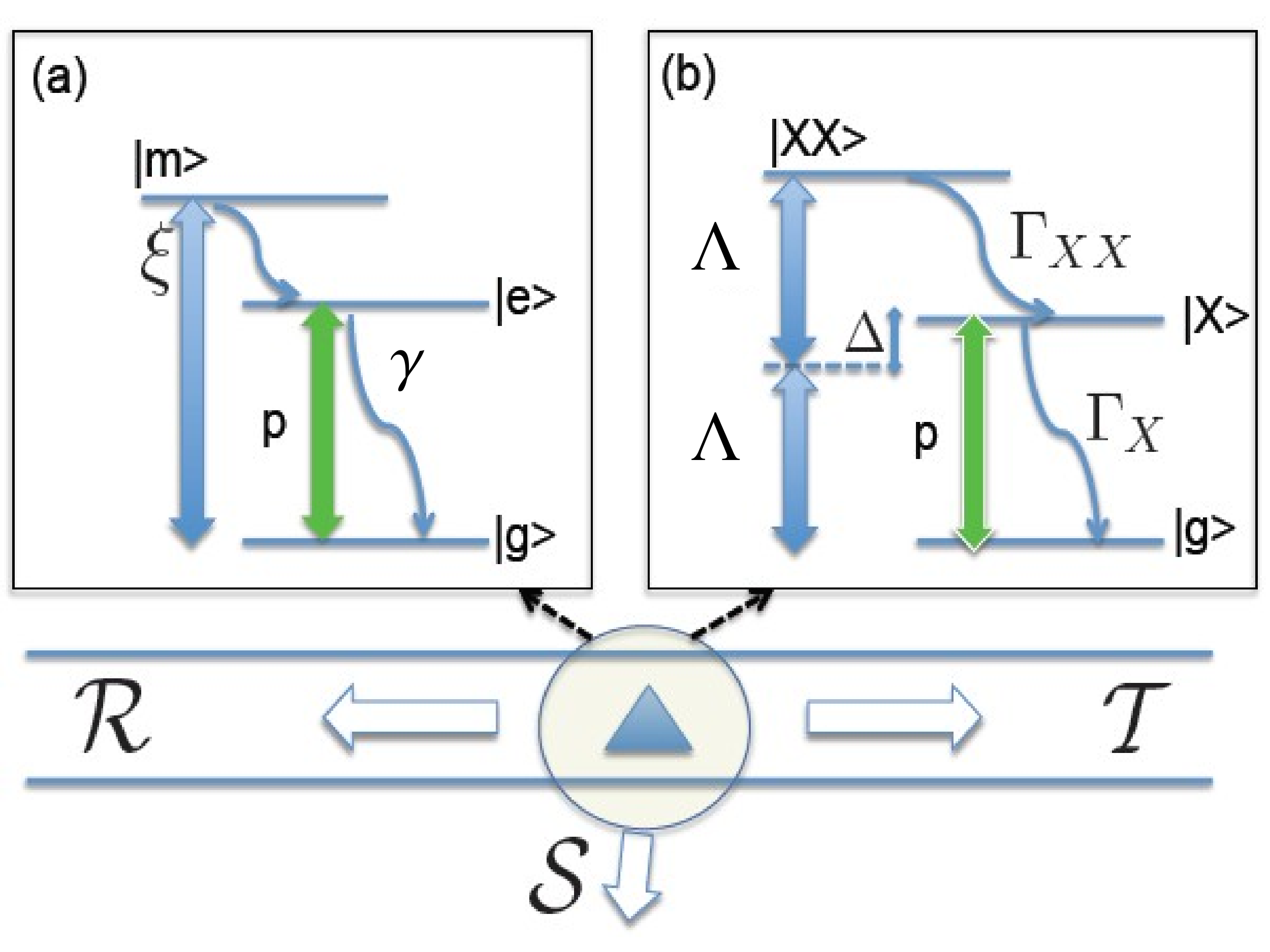}
\caption{Level scheme of the one-D atom under study, (a) under incoherent pumping
or (b) coherent two-photon excitation in the QD case. The notations are introduced in the text.}
\label{fig:scheme}
\end{center}
\end{figure}

The scheme of the two-level emitter of frequency $\omega_A$ interacting with
the continuum of modes inside a 1D waveguide
is pictured in fig.\ref{fig:scheme}(a). The quantized field in the Heisenberg picture writes
$E(z,t) = E^{(+)}(z,t) + E^{(-)}(z,t)$ \cite{glauber}, where
$E^{(+)}(z,t) = i\sum_{\omega} \epsilon_\omega \left\{a_\omega(t)\ e^{i k z}
+ b_\omega(t)\ e^{-i k z} \right\}.$
We have explicitly separated the forward $a_\omega$ from the backward $b_\omega$ propagating modes \cite{domokos}.
The electric field per photon is $\epsilon_\omega$.
The atomic emission is eventually stimulated by a laser of frequency $\omega_L$ injected in the waveguide. This
is well described by a coherent field $\alpha_L$ in the guided mode of the same frequency, the other modes being in the vacuum \cite{cohen}.
The coupling Hamiltonian between the field and the atomic dipole written in the rotating wave approximation is
$H_I = -i\hbar\sum_\omega g_\omega \left[ \sigma_{+}(a_\omega e^{i\frac{\omega}{c} z_A}+b_\omega e^{-i\frac{\omega}{c} z_A}) - \mbox{H.c.} \right],$
where $z_A$ is the position of the atom inside the
waveguide, that we further take equal to $0$. The atomic operators are denoted $\sigma_{+} = |e\rangle\langle g|$, $\sigma_{-}
= \sigma_{+}^\dagger$ and $\sigma_z = (\sigma_{+}\sigma_{-}-\sigma_{-}\sigma_{+})/2$.
The coupling frequency is defined by $ g_\omega = d\epsilon_\omega/\hbar$,
where $d$ is the electric dipole between $|g\rangle$ and $|e\rangle$ states.
 In addition to the Hamiltonian part, an incoherent pump $\xi$ can be added to invert the atomic population.
As pictured in fig.\ref{fig:scheme}, such a mechanism is obtained by resonantly pumping an ancilla level $|m\rangle$ that immediately
decays towards the excited level $\ket{e}$.
We also include a pure dephasing rate $\gamma^*$ \cite{carmichael, alexia}, related to electrostatic fluctuations of the environment \cite{GCassabois},
usually present in solid-state artificial atoms.
The total decay rate is $\gamma = \gamma_0+\gamma_1$,
where $\gamma_1$ is the relaxation rate due to the coupling with the 1D continuum. Unavoidable coupling
to the other modes of the 3D electromagnetic environment gives rise to the additional rate $\gamma_0$.
Time evolution of the operators is given by a set of
coupled Heisenberg-Langevin equations, written in the
frame rotating at the laser frequency in the Markovian approximation \cite{cohen, domokos}.
This leads to the following set of equations, valid in the 1D geometry:

\beq
\frac{d}{dt}  \langle \sigma_{-}\rangle = -  \left(\frac{\gamma + \gamma^*+\xi}{2}-i \delta_L \right) \langle \sigma_{-}\rangle
 + \Omega\ \langle \sigma_{z}\rangle, \nonumber\\
\eeq
\beq
\frac{d}{dt}  \langle \sigma_{z}\rangle = -  (\gamma+\xi) \left(\langle \sigma_{z}\rangle+\frac{1}{2}\right) + \xi
 - \Omega\  \Re[\langle \sigma_{-}\rangle ],
\label{bloch}
\eeq

where $\delta_L = \omega_L - \omega_A$ is the detuning between the atom and the laser
and $\Re$ stands for the real part.
In the following, we will always consider the resonant case $\delta_L=0$.
The Rabi frequency $\Omega$ characterizes the coupling between the atom and the field and equals $\Omega= \gamma \sqrt{2\beta}\sqrt{p}$, where
$p=|\alpha_L|^2/(\pi\rho_{1_D}\gamma)$ is the number of incoming photons per atomic lifetime, $\rho_{1_D} = L/\pi c$ is the density
of modes of the continuum with length of quantization $L$ and $\beta = \gamma_1/(\gamma_0+\gamma_1)$ quantifies the 1D character
of the system. An ideal 1D atom corresponds to $\beta=1$, a limit almost reached in circuit QED \cite{astafiev}. Other experimental setups
also provide nearly ideal 1D systems, namely atoms in strongly dissipative cavity ($\beta = 0.96$ \cite{1D} ), QDs in photonic nanowires ($\beta = 0.95$ \cite{jmg}),
or in photonic crystal waveguides ($\beta = 0.98$ \cite{englund}).

As far as the light field is concerned, we derive the photodetection relation
proper to the 1D geometry, valid for all $z$ and $t>|z|/c$:
\begin{align}
& E^{(+)}(z,t) = E_{a,\mbox{free}}(z,t) + E_{b,\mbox{free}}(z,t)  \nonumber\\
&\qquad + \eta \Big\{ \sigma_{-}\left(t- \frac{z}{c}\right)\Theta(z)
+\sigma_{-}\left(t+\frac{z}{c}\right)\Theta(-z) \Big\},
\label{photodetection}
\end{align}
where we have introduced the parameter $\eta = i\epsilon_{\omega_L}\sqrt{\beta/2}$.
The counterpropagating free field operators are $E_{a,\mbox{free}}(z,t)
= i\sum_\omega \epsilon_\omega a_\omega(0) e^{-i\omega(t-z/c)}$ and
$E_{b,\mbox{free}}(z,t)
= i\sum_\omega \epsilon_\omega b_\omega(0) e^{-i\omega(t+z/c)}$.
Finally, the expressions for the powers $\gamma \langle E^{(-)}E^{(+)}\rangle/\epsilon_{\omega_L}^2$ radiated in the transmission and reflection channels, respectively denoted ${\cal T}$ and ${\cal R}$,
are expressed in number of photons per second and read

\beq
\mathcal{T} = \gamma p + \Omega\ \Re[\langle \sigma_{-}\rangle] + \frac{\gamma\beta}{2}\left(\langle \sigma_z \rangle + \frac{1}{2}\right),
\label{t}\eeq
\beq
\mathcal{R} = \frac{\gamma\beta}{2}\left(\langle \sigma_z \rangle + \frac{1}{2}\right),
\label{r}\eeq

whereas the power dissipated in the leaky modes is denoted ${\cal S}$ and checks
$\mathcal{S}= \gamma(1-\beta)\left(\langle \sigma_z \rangle + \frac{1}{2}\right).$
The first term in ${\cal T}$ equals the incoming laser power $\gamma p$.
The last term is due to spontaneous emission and equally contributes to ${\cal R}$ and ${\cal T}$.
It scales like the excited population $P_e=\langle\sigma_z \rangle+1/2$, so that this atomic observable can be continuously monitored
by observing the reflection ${\cal R}$ or the leaky channel ${\cal S}$.
The interference term $\Omega\ \Re[\langle \sigma_{-}\rangle]$ plays a key role in the 1D geometry under study.
It equals indeed $\mathcal{T}-\gamma p-\mathcal{R}$, allowing to compare the net transmitted
power $\mathcal{T} - \gamma p$ to the reflected power $\mathcal{R}$.
It quantifies then the preferred emission channel. This quantity also
acts on the evolution of the population $P_e$ as it appears in eq.(\ref{bloch}). Following the notations
of a seminal paper by Mollow \cite{mollow}, this term exactly satisfies $\Omega\ \Re[\langle \sigma_{-}\rangle] = -\mathcal{W}[p]$,
where $\mathcal{W}[p]$ stands for the coherent atomic absorption.

\begin{figure}[h!]
\begin{center}
\includegraphics[width=0.8\linewidth]{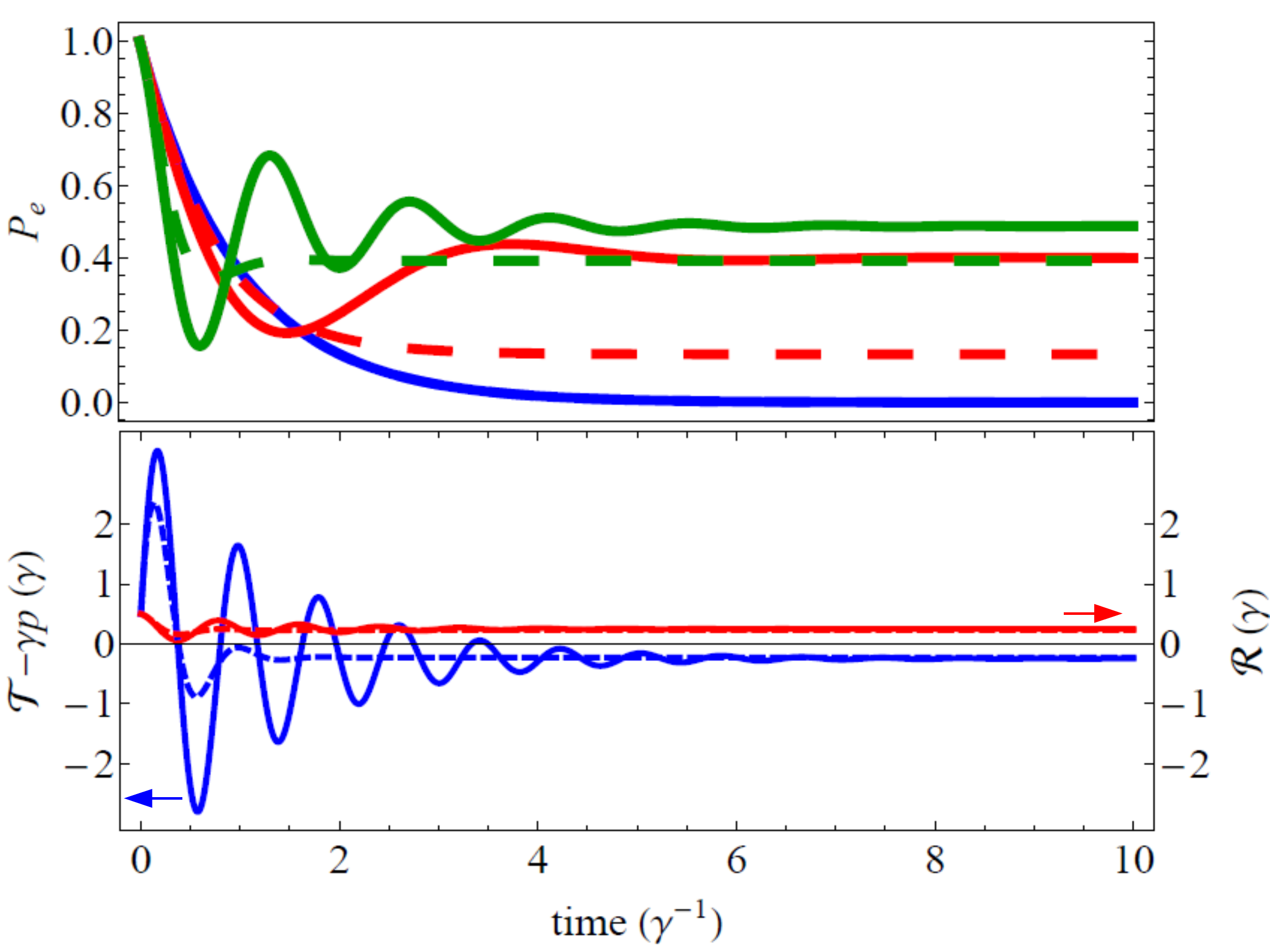}
\caption{(Color online).(Top) Excited state population $P_e(t)$ for $p = 0$ (blue), $p = 1$ (red), $p = 10$ (green). (Bottom) ${\cal T}-\gamma p$ (blue)
and ${\cal R}$(red) for $p = 30$, showing that net transmission overcomes reflection whenever stimulated emission takes place,
in the transient regime. Dashed curves correspond to $\gamma^* = 10\gamma$. For all cases, $\beta = 1$.}
\label{fig:pe1}
\end{center}
\end{figure}

We first consider the transient regime where the incoherent pump is switched off ($\xi=0$), and the atom initially prepared in the excited state $\ket{e}$
is driven by a cw resonant field $p$. The evolution of the population $P_e(t)$ is given by solving eqs.(\ref{bloch}), which correspond to standard Bloch equations
in the case $\xi=0$ under study. It is plotted in Fig.\ref{fig:pe1} with $\beta=1$.
The case $p=0$ corresponds to the damped regime. It is characterized by an exponential decay, typical for the spontaneous emission of a photon into the waveguide.
As it can be seen in the figure, increasing the pump stimulates this emission. However, it appears that stimulated emission does not make the atomic decay
faster, but reversible: this is the non-linear regime of Bloch equations, characterized by the coherent exchange of photons between
the atom and the field (Rabi oscillations) at the rate $\Omega$. This regime is reached when $\Omega$ overcomes the typical dephasing and damping rates $\gamma, \gamma^*$.
When $\gamma^*=0$, this condition boils down to $p \propto \beta^{-1}$ (see eq.\ref{bloch}), which corresponds to $p \sim1$ in the ideal 1D case plotted in the figure.
Therefore, a single photon per lifetime is enough to saturate a 1D atom, as already evidenced in a different context \cite{GNL, kojima}.
If $\beta<1$, a higher pump power will be necessary to reach the non-linear regime. Single photon sensitivity is also altered by pure dephasing, as it appears
in Fig.\ref{fig:pe1} where we have plotted the same quantity with $\gamma^*=10\gamma$, a typical value for quantum dots \cite{alexia}
(note this is an upper bound, pure dephasing rates as low as $\gamma^* = 0.15 \gamma$ being currently reached in circuit QED \cite{astafiev}).
Still, it appears that with realistic parameters, the power needed to reach stimulation remains of the order of few photons per lifetime, so that the giant sensitivity of the device is preserved.
Finally, note that the observed Rabi oscillation is classical and does not lead to any entanglement with the field, contrary to the case of an atom coupled to a monomode cavity \cite{cqed},
another medium showing single photon sensitivity. In that sense, 1D geometry is similar to low quality factor Ramsey zones used in microwave CQED experiments \cite{davidovich}.

It is also interesting to observe the evolution of the radiated fields in the regime of stimulated emission. It is plotted in Fig.\ref{fig:pe1} for $p=30$.
Rabi oscillations are also visible in the reflected and transmitted fields. In particular, one observes that each decrease in $\mathcal{R}$ corresponds to the
stimulated emission of a photon, which feeds the transmission channel. These processes check ${\cal T}-\gamma p>{\cal R}$, confirming that the ``stimulated channel'' ${\cal T}$ is favored.
Note that on the opposite, if the atom is initially prepared in the ground state $\ket{g}$, emission is favored in the reflection channel at the initial time,
a property that can be exploited to develop single photon transistors \cite{sandoghdar,Lukin}.

\begin{figure}[h!]
\begin{center}
\includegraphics[width=0.8\linewidth]{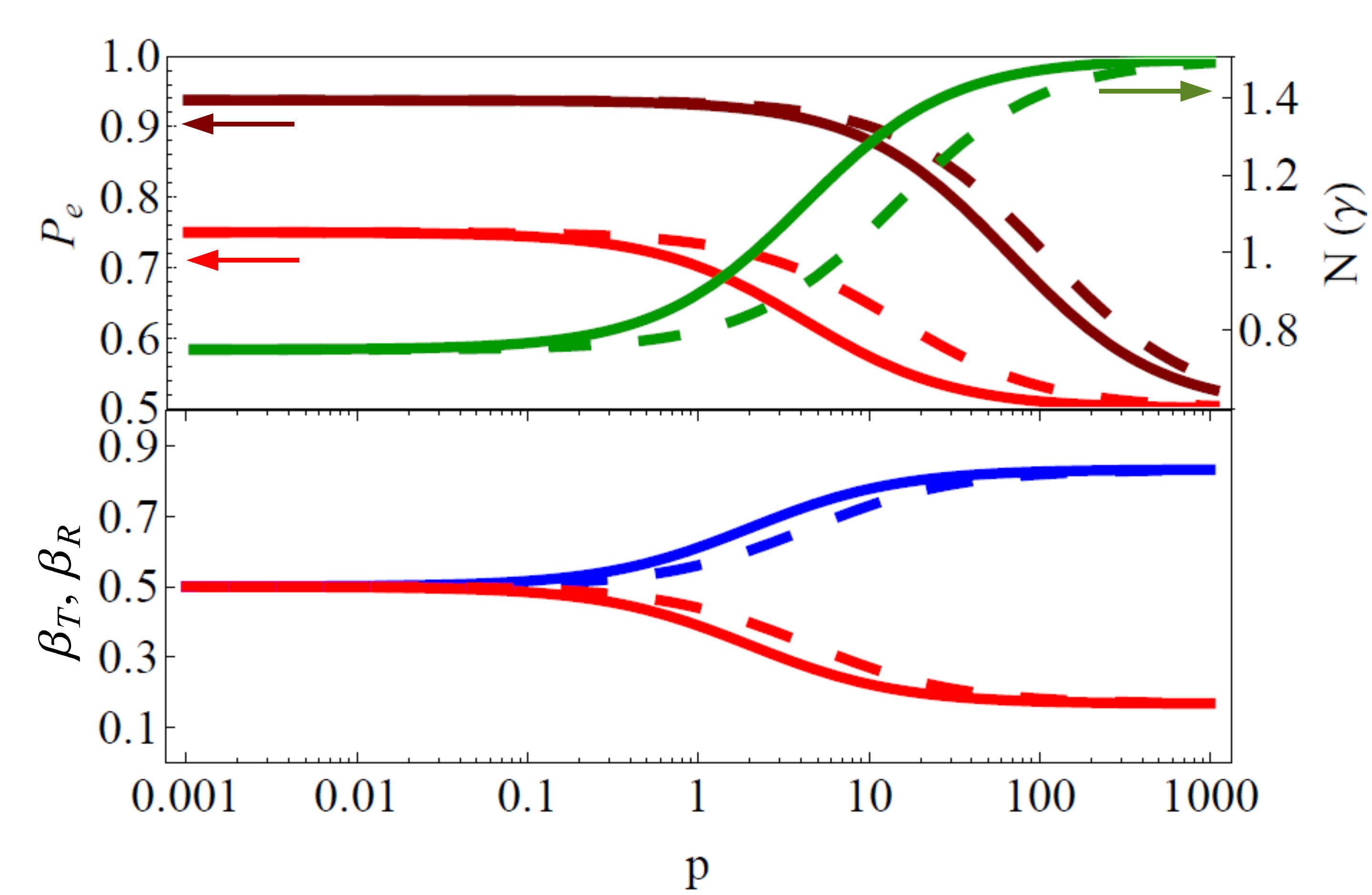}
\caption{(Color Online).(Top) Steady state population (red) of the excited level
(proportional to reflected power) as a function of the resonant pump for $\xi = 3\gamma$(light red) and $15\gamma$(dark red).
Increasing net rate of emitted photons $N$
(green) ranging from $\gamma P_e$ to $\xi/2$ (plotted with $\xi = 3\gamma$).
(Bottom) Ratios  $\beta_T$(blue) and $\beta_R$(red) showing predominance
of emission in the transmission channel for $p > 1$. Dashed curves: $\gamma^* = 10 \gamma$. We took $\beta=1$.}
\label{fig:steadystate}
\end{center}
\end{figure}

Let us now concentrate on the case where the atom is continuously driven
by an incoherent pump $\xi$, and study the influence of the resonant light on the steady
state atomic population $P_{e}$ and radiated fields ${\cal R}$ and ${\cal T}-\gamma p$.
The population is pictured in Fig.\ref{fig:steadystate} as a function of $p$. We have plotted the results for
two different values of the incoherent pump $\xi = 3\gamma$ and $15\gamma$, yielding two different population inversions ($P_e > P_g$) when $p=0$.
We also show the net total rate of photons emitted by the atom $N = \mathcal{T}-\gamma p + \mathcal{R} + \mathcal{S}$.
Moreover, we have defined and plotted the ratios $\beta_R$ ($\beta_T$) of photons
emitted in the reflection (transmission) channel,
in the following way: $\beta_R={\cal R}/N$,  $\beta_T=({\cal T}-\gamma p)/N$.
These quantities measure the propension of the atom to emit in the reflection (transmission)
channels and appear as natural figures of merit for stimulated emission.
Two regimes can be observed in the figure. A vanishing pump $p \rightarrow 0$ gives rise to an incoherent regime characterized by
the spontaneous emission of photons. The excited state population
reads $P_e = \frac{\xi}{\gamma+\xi}$ and the net total rate of emitted photons
is $N = \gamma P_e$. In this regime, no channel is favored, and the net transmitted and reflected fields are equal.
Increasing the pump $p$ to arbitrarily high values sets up the coherent regime of Rabi oscillations. The excited state population $P_e$ decreases,
eventually becoming equal to the ground state population $P_g$, which is the usual limit of Bloch equations when the atom is saturated \cite{cohen}.
Simultaneously, the net total rate of photons increases to $N = \xi/2$. This is an unusual situation where the emitted light power does not
follow the same evolution as the atomic population. As a matter of fact, the rate $N$ represents the rate of photons exchanged between
the atom and the field, which scales like the Rabi frequency $\Omega$ and increases with the pump power $p$.
Simultaneously, the transmission channel is markedly favored with respect to the reflection channel ($\beta_T > \beta_R$).
The transition between these two regimes happens when $p > p_{th}=\frac{(\gamma+\gamma^*+\xi)(\gamma+\xi)}{4\beta \gamma^2}$,
which simplifies in $p_{th} = \frac{1}{4}\left(1+\frac{\xi}{\gamma}\right)^2$ for $\beta = 1$ and $\gamma^*<<\gamma$.
This confirms that Rabi oscillations appear when coherent processes, quantified by $p$, overcome incoherent ones, quantified by $\xi$.
As in the transient case, pure dephasing and lower $\beta$ increase the threshold needed to reach the coherent regime, up to values
that remain of the order of a few photons per lifetime in the physical systems modelled.

\begin{figure}[h!]
\begin{center}
\includegraphics[width=0.8\linewidth]{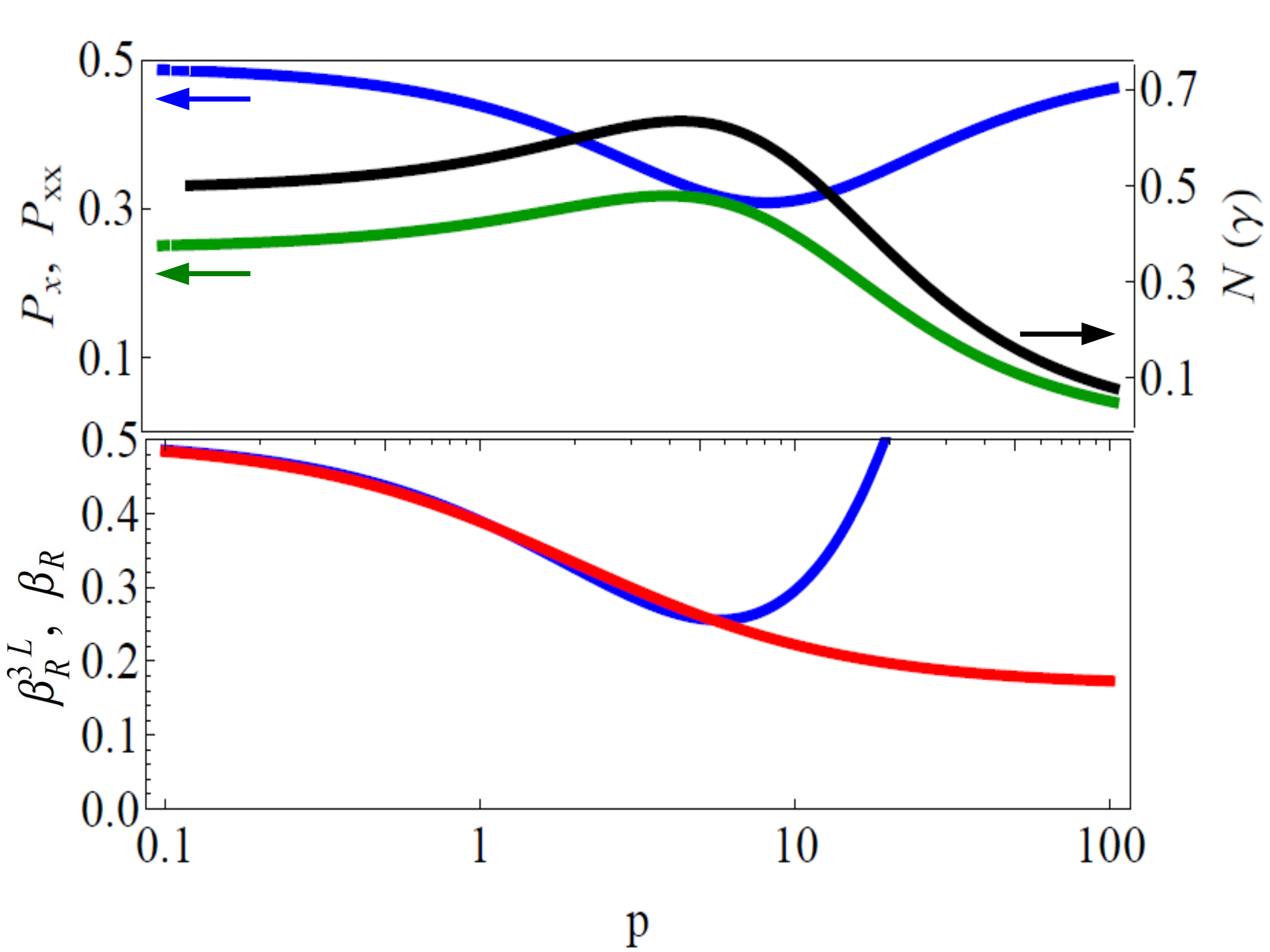}
\caption{(Color online).(Top) Populations $P_X$(blue) and $P_{XX}$(green) of the 3 levels system {\it vs.} p.
Net rate of photons $N$ (black) in the exciton transition.
(Bottom) Comparison between the ratios of emission given by the two ($\beta_R$, red) and three ($\beta^{3L}_R$, blue) levels,
 assuming $\xi\sim 3\gamma$ and $\Lambda^2/\Delta \sim 4\Gamma_X$. In both cases, $\gamma^* = 0$ and $\beta=1$.}
\label{fig:3levels}
\end{center}
\end{figure}

Measuring the ratios $\beta_R$ and $\beta_T$ is experimentally quite demanding.
As a matter of fact, it requires the ability to quantify the total power radiated by the atom, in particular the net transmitted power
${\cal T}-\gamma p$, hence to filter the pump to extract a tiny atomic emission. Therefore, we propose an experimentally feasible way to measure
this quantity, by exploiting a third atomic level $|XX\rangle$ as pictured in fig.\ref{fig:scheme}(b).
This three level structure can model the biexcitonic and the excitonic transitions of a
quantum dot, a terminology that we shall use from now on without losing the generality of the scheme.
Population inversion on the excitonic transition ($P_X > P_g$) is reached by resonantly pumping a biexciton in the dot using two-photon absorption
technique. This mechanism can be described by an effective Hamiltonian $H_{2ph}=\hbar\Lambda^2/\Delta (\ket{XX}\bra{g}\ e^{-i\nu t}+\mbox{h.c.})$,
where $\Delta=(E_{XX}-E_X)/2$ and $\Lambda$ is the Rabi frequency of the pump \cite{twophotons}.
As usual, Lindbladians describe the decays $|XX\rangle \rightarrow |X\rangle $, with rate $\Gamma_{XX}$, and $|X\rangle \rightarrow |g\rangle$, with rate $\Gamma_X$.
The populations of the excitonic $P_X$ and biexcitonic states $P_{XX}$
are computed in the steady state regime, as a function of the resonant probe $p$.
The results are plotted in fig.\ref{fig:3levels}. When $p=0$, the presence of a large pump $\Lambda^2/\Delta > \Gamma_{XX}$ leads to equalize the populations of the ground and biexcitonic state, whereas detailed balance conditions require $P_X/P_{XX} = \Gamma_{XX}/\Gamma_X$. Usual quantum dot parameters satisfy $\Gamma_{XX}\approx 2\Gamma_{X}$ \cite{jmg}, leading to $P_{XX}=P_{g}=1/4$ and $P_X=1/2$.
Increasing the probe $p$  leads to the depletion of the excitonic level because of stimulated emission,
thus to the increase of the steady state biexcitonic population as it appears in fig.\ref{fig:3levels}.
This increase can be monitored by measuring the rate of photons $\Gamma_{XX}\ P_{XX}$ emitted at the biexcitonic frequency,
which provides an easily observable signature of stimulated emission at the single-photon level.
Moreover, we have verified that $\Gamma_{XX}\ P_{XX} = \Gamma_X\ P_X - \mathcal{W}[p] = N$, where ${\cal W}[p]$ is the interference term between the probe and the light emitted at the excitonic frequency, as defined above. Namely, the rate of photons emitted in the biexcitonic line exactly equals the rate $N$ emitted in the excitonic one, taking into account stimulated processes.
Stimulated emission of the excitonic transition can thus be simply monitored, by measuring the rate of photons emitted at the biexcitonic frequency.
This rate can be used to build the ratio $\beta_R$ defined above, without having to measure the net transmitted power.  This is also represented
in fig.\ref{fig:3levels}, where we have plotted $\beta_R^{3L} = \frac{\frac{\Gamma_X}{2}\ P_X}{\Gamma_{XX}\ P_{XX}}$,
the index $3L$ standing for 3 Levels. For the sake of comparison,
we have plotted on the same figure the quantity $\beta_R$ defined in the case of a two-level atom.
The equivalence between the models is clearly shown in the coincidence of both curves for
low power $p$ ($\Omega= \gamma \sqrt{\beta}\sqrt{2p} < \Lambda^2/\Delta$).
A divergence becomes unavoidable when $p$ is strong enough to generate Autler-Townes splitting \cite{autlertownes} of the ground level.
The biexcitonic transition becomes thus out of resonance with the $\Lambda$-driving field. So, the population
of the biexciton state drastically decreases, making the ratio $\beta_R^{3L}$ arbitrarily large and equalizing exciton and
ground state populations.
A possible drawback of experiments performed with quantum dots can be an imperfect two-photon absorption,
leading to an incoherent feeding of the excitonic level {\it via} phonons, even for large $\Delta$ \cite{brunner}.
However, a recent experimental work \cite{flis} shows that the two-photon transition can be made
very clean, so that the incoherent exciton pumping is negligible in this case. Finally,
note that a scheme to fully protect entanglement has been proposed using the same mechanism of biexcitonic pumping and read outs of the light emitted in each possible transition \cite{marcelo}.

In conclusion, we have evidenced signatures of stimulated emission
at the single-photon level, giving rise to potentially observable effects
with state of the art solid-state atoms interacting with 1D light fields.
In particular, we propose a new experiment to probe the stimulated (optical) transition,
based on the monitoring of an ancillary transition.
Properties of 1D atoms evidenced in this work
may be exploited to implement fundamental quantum tasks,
such as single photon optimal cloning, or single photon amplification.

This work was supported by the
Nanosciences Foundation of Grenoble, the CNPq, the Fapemig, and the ANR project ``CAFE''.
The authors thank Lucien Besombes
and Claire Le Gall for the fruitful discussions, and
the Centre for Quantum Technologies in Singapore for its kind hospitality.


\begin{thebibliography}{10}

\bibitem{cqed} M. Brune et al, Phys. Rev. Lett. \textbf{76}, 1800 (1996);

\bibitem{kimble} A. Boca et al, Phys. Rev. Lett. \textbf{93}, 233603 (2004);

\bibitem{1D} Q. A. Turchette et al, Phys. Rev. Lett. \textbf{75}, 4710 (1995); Q.A. Turchette, R.J. Thompson and H.J. Kimble, Appl. Phys. B \textbf{60} S1 (1995);

\bibitem{jmg} J. Claudon et al, Nature Photonics \textbf{4},174 (2010);

\bibitem{englund} D. Englund et al, Nature \textbf{450}, 857 (2007);

\bibitem{Lukin} D. E. Chang, A. S. Sorensen, E. A. Demler and M. D. Lukin, Nature Physics \textbf{3}, 807 (2007);

\bibitem{circuitqed} R. J. Schoelkopf, S. M. Girvin, Nature \textbf{451}, 664 (2008);

\bibitem{astafiev} O. Astafiev et al, Science \textbf{840}, 327 (2010);

\bibitem{atoms} S. A.  Aljunid et al, Phys. Rev. Lett. \textbf{103}, 153601 (2009), M. Stobinska, G. Alber and G. Leuchs,  Eur. Phys. Lett. \textbf{86} 14007 (2009);

\bibitem{sandoghdar} G. Wrigge at al, Nature Physics \textbf{4}, 60 (2008); J. Hwang et al, Nature \textbf{460}, 76 (2009);

\bibitem{GNL} A. Auff\`eves-Garnier et al, Phys. Rev. A \textbf{75}, 053823 (2007);

\bibitem{kojima} K. Kojima,H. F. Hofmann ,S. Takeuchi, and K. Sasaki, Phys. Rev. A \textbf{70}, 013810 (2004);

\bibitem{einstein} A. Einstein, Phys. Z. 18, 121 (1917);

\bibitem{simon} C. Simon et al, Phys. Rev. Lett. \textbf{84}, 2993 (2000);

\bibitem{maser} G. Rempe et al, Phys. Rev. Lett. \textbf{64}, 2783 (1990);

\bibitem{glauber} R. J. Glauber, Phys. Rev. \textbf{130}, 2529 (1963);

\bibitem{domokos} P. Domokos et al, Phys. Rev. A \textbf{65}, 033832 (2002);

\bibitem{cohen} C. Cohen-Tannoudji et al, Photon-Atom interactions, 2ed (2004);

\bibitem{carmichael} H. Carmichael, {\it An Open System Approach to Quantum
Optics}, Springer-Verlag (1993);

\bibitem{alexia} A. Auff\`eves et al, Phys. Rev. A \textbf{79}, 053838 (2009);

\bibitem{GCassabois} I. Favero, et al, Phys. Rev. B \textbf{75}, 073308 (2007);

\bibitem{mollow} B. R. Mollow, Phys. Rev. A \textbf{5}, 2217 (1972);

\bibitem{davidovich} J. I. Kim et al, Phys. Rev. Lett. \textbf{82}, 4737 (1999);

\bibitem{twophotons} M. Brune et al, Phys. Rev. A \textbf{35}, 154 (1987),
E. del Valle et al, Phys. Rev. B \textbf{81}, 035302 (2010);

\bibitem{autlertownes}S. H. Autler, C. H. Townes, Phys. Rev. \textbf{100}, 703 (1955);

\bibitem{brunner} K. Brunner et al, Phys. Rev. Lett. \textbf{73}, 1138 (1994),
     M. Winger et al, Phys. Rev. Lett. \textbf{103}, 207403 (2009);

\bibitem{flis} T. Flissikowski et al, Phys. Rev. Lett. \textbf{92}, 227401 (2004).

\bibitem{marcelo} A.R.R. Carvalho and M. F. Santos, New J. Phys. \textbf{13} 013010 (2011);

\end{thebibliography}
\end{document}